\documentclass[prb,twocolumn]{revtex4}
\usepackage{amsmath,amsfonts}
\usepackage{graphicx}
\usepackage{dcolumn}
\newcolumntype{.}{D{.}{.}{-1}}
\newcommand{\bfr}{\mathbf{r}}

\begin{document}

\title{Improved method for generating exchange-correlation potentials
from electronic wave functions}

\author{Egor Ospadov$^{1}$}
\author{Ilya G. Ryabinkin$^{2}$}
\author{Viktor N. Staroverov$^{1}$}

\affiliation{$^{1}$Department of Chemistry, The University of Western
Ontario, London, Ontario N6A 5B7, Canada}

\affiliation{$^{2}$Department of Physical and Environmental Sciences,
University of Toronto Scarborough, Toronto, Ontario M1C 1A4, Canada}

\date{\today}

\begin{abstract}

Ryabinkin, Kohut, and Staroverov (RKS) [Phys.~Rev.~Lett.~\textbf{115},
083001 (2015)] devised an iterative method for reducing many-electron
wave functions to Kohn--Sham exchange-correlation potentials,
$v_\text{XC}(\bfr)$. For a given type of wave function, the RKS
method is exact (Kohn--Sham-compliant) in the basis-set limit; in a
finite basis set, it produces an approximation to the corresponding
basis-set-limit $v_\text{XC}(\bfr)$. The original RKS procedure works
very well for large basis sets but sometimes fails for commonly used
(small and medium) sets. We derive a modification of the method's working
equation that makes the RKS procedure robust for all Gaussian basis
sets and increases the accuracy of the resulting exchange-correlation
potentials with respect to the basis-set limit.

\end{abstract}

\maketitle

\section{Introduction}

Recently, the present authors and their
co-workers\cite{Ryabinkin.2015.PRL.115.083001,Cuevas-Saavedra.2015.JCP.143.244116,%
Cuevas-Saavedra.2016.MP.114.1050,Kohut.2016.PCCP.18.20938,%
Ryabinkin.2016.JCP.145.037102} developed a method for
constructing Kohn--Sham (KS) exchange-correlation
potentials, $v_\text{XC}(\bfr)$, from electronic wave
functions for nondegenerate ground states that are pure-state
$v$-representable.\cite{Levy.1982.PRA.26.1200,Englisch.1983.P.121A.253}
In this method, $v_\text{XC}(\bfr)$ is generated 
by iterating an analytic expression that relates this potential
to the interacting two-electron reduced density matrix (2-RDM)
of the system. Refs.~\citenum{Ryabinkin.2015.PRL.115.083001}
and \citenum{Cuevas-Saavedra.2015.JCP.143.244116} describe
two implementations of our technique based on slightly different but
equivalent expressions for $v_\text{XC}(\bfr)$,
Ref.~\citenum{Cuevas-Saavedra.2016.MP.114.1050} presents
a general approach for deriving such expressions,
whereas Refs.~\citenum{Kohut.2016.PCCP.18.20938}
and \citenum{Ryabinkin.2016.JCP.145.037102}
elaborate on the implications. Since the two published
variants\cite{Ryabinkin.2015.PRL.115.083001,Cuevas-Saavedra.2015.JCP.143.244116}
of our method are interchangeable, we will refer to them collectively as
the Ryabinkin--Kohut--Staroverov (RKS) procedure, after the authors
of Ref.~\citenum{Ryabinkin.2015.PRL.115.083001}. In the special case
of Hartree--Fock (HF) wave functions, the RKS procedure reduces
to the method of Refs.\citenum{Ryabinkin.2013.PRL.111.013001} and
\citenum{Kohut.2014.JCP.140.18A535}.

The RKS method is \textit{not} a KS inversion technique, that is, it does
not focus on finding the KS potential that reproduces a given \textit{ab
initio} electron density $\rho^\text{WF}(\bfr)$. The KS inversion problem
is ill-conditioned\cite{Savin.2003.IJQC.93.166} and its solution is
not unique when the KS equations are solved in a finite one-electron
basis set.\cite{Harriman.1983.PRA.27.632,Gorling.1995.PRA.51.4501,%
Staroverov.2006.JCP.124.141103,Pino.2009.TCA.123.189} The
objective of the RKS method is to approximate the
basis-set-limit $v_\text{XC}(\bfr)$ of the system when both
wave-function and KS calculations are done using a finite
basis set. RKS potentials are obtained from the 2-RDM via an
analytic expression for $v_\text{XC}(\bfr)$ that is exact in a
complete (infinite) basis set but not in a finite one. As
a consequence, they are unambiguous and uniform, but the density
$\rho^\text{KS}(\bfr)$ generated by an RKS potential is exactly equal
to $\rho^\text{WF}(\bfr)$ only in the basis-set limit. This is to be
contrasted with KS inversion techniques,\cite{Gorling.1992.PRA.46.3753,%
Wang.1993.PRA.47.R1591,Zhao.1994.PRA.50.2138,Leeuwen.1994.PRA.49.2421,%
Schipper.1997.TCA.98.16,Peirs.2003.PRA.67.012505,Wu.2003.JCP.118.2498,%
Ryabinkin.2012.JCP.137.164113,Hollins.2017.JPCM.29.04LT01} where
the requirement that $\rho^\text{KS}(\bfr)$ match $\rho^\text{WF}(\bfr)$ in
\textit{any} basis set can result in potentials that oscillate,
diverge, and look nothing like the $v_\text{XC}(\bfr)$
of the basis-set limit $\rho^\text{WF}(\bfr)$ for the same
system.\cite{Schipper.1997.TCA.98.16,Mura.1997.JCP.106.9659} Thus, KS
inversion and RKS methods pose different questions and give different
answers in finite basis sets.

In our experience, the RKS procedure works best for large
uncontracted basis sets such as the universal Gaussian basis
set (UGBS).\cite{Castro.1998.JCP.108.5225} For general-purpose
basis sets such as cc-pVXZ,\cite{Dunning.1989.JCP.90.1007}
cc-pCVXZ,\cite{Woon.1995.JCP.103.4572} and 6-311G*, it often works
well, but sometimes produces deformed potentials or even fails to
converge (see examples below). Here we propose a modification to the
RKS method that eliminates all such problems, increases the uniformity
of potentials obtained in various Gaussian basis sets, and substantially
improves the accuracy of potentials generated in small basis sets
with respect to the basis-set limit.

\section{RKS method and its modification}

The exact expression for $v_\text{XC}$
that lies at the heart of the RKS method was
obtained\cite{Ryabinkin.2015.PRL.115.083001,Cuevas-Saavedra.2015.JCP.143.244116}
by combining two local energy balance equations derived within the KS and
\textit{ab initio} wave-function formalisms for a given $N$-electron
system. These two equations contain the molecular electrostatic
potential but differ in all other terms.
The fact that both equations describe the same system is expressed
by the condition
\begin{equation}
 \rho^\text{KS}(\bfr) = \rho^\text{WF}(\bfr). \label{eq:rhos}
\end{equation}
When one local energy balance equation is subtracted from the other,
the electrostatic potential drops out
and we obtain the following intermediate result:
\begin{equation}
 v_\text{XC} = v_\text{XC}^\text{hole}
  + \bar{\epsilon}^\text{KS} - \bar{\epsilon}^\text{WF}
  + \frac{\tau_L^\text{WF}}{\rho^\text{WF}} 
 - \frac{\tau_L^\text{KS}}{\rho^\text{KS}},
  \label{eq:vxc-1}
\end{equation}
where each quantity is a function of $\bfr$. Here
\begin{equation}
 v_\text{XC}^\text{hole}(\bfr)
  = \int \frac{\rho_\text{XC}(\bfr,\bfr_2)}{|\bfr-\bfr_2|}\,d\bfr_2
\end{equation}
is the potential of the exchange-correlation hole charge\cite{Parr.1989}
derived from the interacting 2-RDM,
\begin{equation}
 \bar{\epsilon}^\text{KS} = \frac{1}{\rho^\text{KS}}
 \sum_{i=1}^N \epsilon_i |\phi_i|^2   \label{eq:eps-KS}
\end{equation}
is the average local KS orbital energy,
in which $\phi_i$ are the spatial parts of KS spin-orbitals,
$\epsilon_i$ are their eigenvalues, and
\begin{equation}
 \rho^\text{KS} = \sum_{i=1}^N |\phi_i|^2.
\end{equation}
The next quantity, defined by
\begin{equation}
 \bar{\epsilon}^\text{WF} = \frac{1}{\rho^\text{WF}}
 \sum_{j} \lambda_j|f_j|^2, \label{eq:eps-WF}
\end{equation}
is the \textit{ab initio} average local electron
energy,\cite{Ryabinkin.2014.JCP.141.084107,Kohut.2016.JCP.145.074113}
in which $f_j$ are the eigenfunctions of the generalized Fock operator,
$\lambda_j$ are their eigenvalues, and $\rho^\text{WF}$ is the
\textit{ab initio} electron density. The summation in Eq.~\eqref{eq:eps-WF}
extends over all eigenfunctions $f_j$ whose number is equal to the number of
one-electron basis-set functions. We choose to write the \textit{ab initio}
electron density as 
\begin{equation}
 \rho^\text{WF} = \sum_{j} n_j |\chi_j|^2,
\end{equation}
where $\chi_j$ are the natural orbitals and $n_j$ are their
occupation numbers. The remaining quantities are
\begin{equation}
 \tau_L^\text{WF} = -\frac{1}{2} \Re \left[
 \sum_j n_j \chi_j^* \nabla^2 \chi_j \right],
\end{equation}
the Laplacian form of the interacting (\textit{ab initio})
kinetic-energy density expressed through natural orbitals, and
\begin{equation}
 \tau_L^\text{KS} = -\frac{1}{2} \Re \left[ \sum_{i=1}^N
   \phi_i^{*}\nabla^2\phi_i \right],
\end{equation}
the Laplacian form of the noninteracting (KS) kinetic-energy
density. Note that Eq.~\eqref{eq:vxc-1} is one of an entire class
of exact expressions for $v_\text{XC}$.\cite{Cuevas-Saavedra.2016.MP.114.1050,
Buijse.1989.PRA.40.4190,Baerends.1997.JPCA.101.5383,Chong.2002.JCP.116.1760}

For reasons discussed below, the RKS procedure uses \textit{not}
Eq.~\eqref{eq:vxc-1} but a different expression obtained
from Eq.~\eqref{eq:vxc-1} by applying to $\tau_L^\text{WF}$ and
$\tau_L^\text{KS}$ the identity
\begin{equation}
 \tau_L = \tau - \frac{1}{4} \nabla^2 \rho,  \label{eq:tauL}
\end{equation}
where $\tau$ denotes the respective positive-definite form of the kinetic-energy
density. The terms $\nabla^2\rho^\text{WF}/4\rho^\text{WF}$
and $\nabla^2\rho^\text{KS}/4\rho^\text{KS}$ cancel out because of
Eq.~\eqref{eq:rhos}, and Eq.~\eqref{eq:vxc-1} becomes
\begin{equation}
 v_\text{XC} = v_\text{XC}^\text{hole}
  + \bar{\epsilon}^\text{KS} - \bar{\epsilon}^\text{WF}
  + \frac{\tau^\text{WF}}{\rho^\text{WF}} - \frac{\tau^\text{KS}}{\rho^\text{KS}},
  \label{eq:vxc-2}
\end{equation}
where
\begin{equation}
 \tau^\text{WF} = \frac{1}{2} \sum_{j} n_j |\nabla\chi_j|^2
\end{equation}
and
\begin{equation}
 \tau^\text{KS}= \frac{1}{2} \sum_{i=1}^N |\nabla\phi_i|^2.
\end{equation}

Equation~\eqref{eq:vxc-2} is the basis of the RKS method.  To construct
$v_\text{XC}$ by this technique one needs to compute all of the
terms on the right-hand side of Eq.~\eqref{eq:vxc-2}. The terms
$v_\text{XC}^\text{hole}$, $\tau^\text{WF}$, and $\rho^\text{WF}$
are extracted from an \textit{ab initio} wave function, but
$\bar{\epsilon}^\text{KS}$ and $\tau^\text{KS}$ are initially unknown
because they depend on $\phi_i$ and $\epsilon_i$, which in turn depend
on $v_\text{XC}$. In Refs.~\onlinecite{Ryabinkin.2015.PRL.115.083001}
and \onlinecite{Cuevas-Saavedra.2015.JCP.143.244116}, we showed that
it is possible to simultaneously solve for $v_\text{XC}$ and the associated
KS orbitals by starting with a reasonable initial guess for
$\phi_i$ and $\epsilon_i$ and iterating Eq.~\eqref{eq:vxc-2} via
the KS equations until the potential $v_\text{XC}$ becomes self-consistent.
In a finite basis set, this potential is such that
$\rho^\text{KS}\neq\rho^\text{WF}$ even at convergence.

Equations~\eqref{eq:vxc-1} and \eqref{eq:vxc-2} are both exact (KS-compliant)
only when all their right-hand-side ingredients are obtained in a
complete basis set. This is because the two local energy
balance equations leading to Eq.~\eqref{eq:vxc-1} were derived
by analytically inverting the KS and generalized Fock eigenvalue
problems,\cite{Ryabinkin.2015.PRL.115.083001,Cuevas-Saavedra.2015.JCP.143.244116}
and analytic inversion of operator eigenvalue
problems amounts to employing a complete basis
set. Refs.~\citenum{Schipper.1997.TCA.98.16},
\citenum{Mura.1997.JCP.106.9659}, and
\citenum{Gaiduk.2013.JCTC.9.3959} demonstrate the dramatic
effect of basis-set incompleteness on the inverted KS equation,
whereas Refs.~\citenum{Ryabinkin.2016.JCP.145.037102} and
\citenum{Kohut.2016.JCP.145.074113} illustrate it for the generalized
Fock eigenvalue problem.

In a finite basis set, Eqs.~\eqref{eq:vxc-1} and \eqref{eq:vxc-2}
are not even equivalent because Eq.~\eqref{eq:rhos}, which
links them, does not hold from the start of iterations.
Previously we found that iterations of Eq.~\eqref{eq:vxc-1}
hardly ever converge, whereas iterations of Eq.~\eqref{eq:vxc-2}
converge for many, but not all, standard Gaussian basis
sets. We now argue that Eq.~\eqref{eq:vxc-2} works better than
Eq.~\eqref{eq:vxc-1} because in Eq.~\eqref{eq:vxc-2} the difference
$\nabla^2\rho^\text{WF}/4\rho^\text{WF}-\nabla^2\rho^\text{KS}/4\rho^\text{KS}$
is set to its basis-set-limit value of zero even when
$\rho^\text{KS}\neq\rho^\text{WF}$, so the resulting
finite-basis-set $v_\text{XC}$ can get closer to the basis-set-limit potential.
Motivated by this idea, we propose the following
improvement upon Eq.~\eqref{eq:vxc-2}.

Let us assume for simplicity that all $\phi_i$ are real.
Using the Lagrange identity\cite{Mitrinovic.1970} we write
\begin{align}
 2\rho^\text{KS}\tau^\text{KS} & =
 \left( \sum_{i=1}^N \phi_i^2 \right) \left(
 \sum_{i=1}^N |\nabla\phi_i|^2 \right) \label{eq:Lagrange} \\
 & = \left| \sum_{i=1}^N \phi_i\nabla\phi_i \right|^2
 + \sum_{i<j}^N |\phi_i\nabla\phi_j - \phi_j \nabla\phi_i|^2. \nonumber
\end{align}
Recognizing that $|\sum_{i=1}^N \phi_i\nabla\phi_i|^2 = |\nabla\rho^\text{KS}|^2/4$
and dividing Eq.~\eqref{eq:Lagrange} through by $2\rho^\text{KS}$ we have
(cf.~Ref.~\citenum{Tal.1978.IJQCS.12.153})
\begin{equation}
  \tau^\text{KS} = \tau_W^\text{KS} + \tau_P^\text{KS}, \label{eq:tauKS-P}
\end{equation}
where $\tau_W^\text{KS}=|\nabla\rho^\text{KS}|^2/8\rho^\text{KS}$
is the von Weizs\"{a}cker noninteracting kinetic-energy density and
\begin{equation}
  \tau_P^\text{KS} = \frac{1}{2\rho^\text{KS}}
  \sum_{i<j}^N |\phi_i\nabla\phi_j - \phi_j \nabla\phi_i|^2
\end{equation}
is a quantity which we call the Pauli kinetic-energy density
(the name is motivated by Ref.\citenum{Levy.1988.PRA.38.625}).
Similarly, assuming real natural orbitals and applying the Lagrange identity
to the product $2\rho^\text{WF}\tau^\text{WF}$ we obtain
\begin{equation}
  \tau^\text{WF} = \tau_W^\text{WF} + \tau_P^\text{WF}, \label{eq:tauWF-P}
\end{equation}
where $\tau_W^\text{WF}=|\nabla\rho^\text{WF}|^2/8\rho^\text{WF}$ and
\begin{equation}
  \tau_P^\text{WF} = \frac{1}{2\rho^\text{WF}}
   \sum_{i<j} n_i n_j |\chi_i\nabla\chi_j - \chi_j \nabla\chi_i|^2.
\end{equation}
Next we substitute Eqs.~\eqref{eq:tauKS-P} and \eqref{eq:tauWF-P} into
Eq.~\eqref{eq:vxc-2}. In view of Eq.~\eqref{eq:rhos},
the terms $\tau_W^\text{KS}/\rho^\text{KS}$ and
$\tau_W^\text{WF}/\rho^\text{WF}$ cancel out and we arrive
at the following new expression,
\begin{equation}
 v_\text{XC} = v_\text{XC}^\text{hole}
  + \bar{\epsilon}^\text{KS} - \bar{\epsilon}^\text{WF}
  + \frac{\tau_P^\text{WF}}{\rho^\text{WF}} - \frac{\tau_P^\text{KS}}{\rho^\text{KS}},
  \label{eq:vxc-3}
\end{equation}
which is the main result of this work. Just like
Eqs.~\eqref{eq:vxc-1} and \eqref{eq:vxc-2}, Eq.~\eqref{eq:vxc-3}
is KS-compliant only in the basis-set-limit. In
a finite basis set, it should give a better
approximation to the basis-set-limit $v_\text{XC}$
than Eq.~\eqref{eq:vxc-2} because it sets the quantity
$\tau_W^\text{WF}/\rho^\text{WF}-\tau_W^\text{KS}/\rho^\text{KS}$
to its basis-set-limit value of zero even when $\rho^\text{KS}\neq\rho^\text{WF}$.
We will refer to the variant of our method using Eq.~\eqref{eq:vxc-3} as the modified
RKS (mRKS) procedure. 

The mRKS procedure is exactly the same as the
original RKS method\cite{Ryabinkin.2015.PRL.115.083001,%
Cuevas-Saavedra.2015.JCP.143.244116,Kohut.2016.PCCP.18.20938}
except that the former uses Eq.~\eqref{eq:vxc-3} in place of
Eq.~\eqref{eq:vxc-2}. Therefore, we will not describe the mRKS algorithm
in detail here but only emphasize the following important points. The
equality $\rho^\text{KS}=\rho^\text{WF}$ plays a key role in the
derivation of Eqs.~\eqref{eq:vxc-2} and \eqref{eq:vxc-3}, but it is
not imposed when these equations are solved by iteration. Thus, there is no
such thing as a ``target density" in the RKS and mRKS methods, and the extent
to which $\rho^\text{KS}$ deviates from $\rho^\text{WF}$ at convergence
is controlled implicitly through the choice of one-electron basis set.
For internal consistency, the RKS and mRKS procedures use the same
one-electron basis set to generate the \textit{ab initio} wave function
and to solve the KS equations in the iterative part of the algorithm.
The Hartree (Coulomb) contribution to the KS Hamiltonian matrix is
always computed using $\rho^\text{KS}$ (not $\rho^\text{WF}$); we do it
analytically in terms of Gaussian basis functions. Matrix elements of
$v_\text{XC}$ are evaluated using saturated Gauss--Legendre--Lebedev
numerical integration grids. We consider $v_\text{XC}$ converged
when the difference between two consecutive KS density matrices drops
below 10$^{-10}$ in the root-mean-square sense. Both the original and
modified RKS procedures require direct inversion of the iterative
subspace\cite{Pulay.1982.JCC.3.556} to converge the potential in
self-consistent-field (SCF) iterations; the mRKS procedure typically
takes one or two dozen iterations, RKS up to a few dozen. The converged
$v_\text{XC}$ is independent of the initial guess; KS orbitals and
orbital energies from any standard density-functional approximation
are adequate as a starting point for systems with a single-reference
character.  For this work, we re-implemented the RKS and mRKS methods
by modifying the SCF and multiconfigurational SCF links of a more recent
version of the \textsc{gaussian 09} program.\cite{G09.E1-short-JCP}

\section{Comparison of the original and modified RKS methods}
\label{sec:results}

To demonstrate the practical advantages of Eq.~\eqref{eq:vxc-3} over
Eq.~\eqref{eq:vxc-2} we compared exchange-correlated potentials
generated by the mRKS and RKS methods from various atomic and
molecular \textit{ab initio} wave functions. The wave functions
were of three types: HF, complete active space SCF (CASSCF), and
full configuration interaction (FCI). Wave functions of each type
were obtained using a series of standard Gaussian one-electron
basis sets varying between minimal (STO-3G) and very large
(UGBS). All basis sets were taken from the Basis Set Exchange
Database.\cite{Feller.1996.JCC.17.1571,Schuchardt.2007.JCIM.47.1045}

For each wave function, we report three relevant properties:
the total interacting kinetic energy
\begin{equation}
 T = - \frac{1}{2} \sum_{j} n_j \langle \chi_j | \nabla^2 | \chi_j \rangle,
  \label{eq:T}
\end{equation}
the \textit{ab initio} exchange-correlation energy
\begin{equation}
 E_\text{XC}^\text{WF} = \frac{1}{2} \int \rho^\text{WF}(\bfr)
  v_\text{XC}^\text{hole}(\bfr)\,d\bfr,
  \label{eq:EXC-WF}
\end{equation}
and the first ionization energy extracted
from the wave function by the extended Koopmans
theorem\cite{Day.1974.IJQCS.8.501,Smith.1975.JCP.62.113,%
Day.1975.JCP.62.115,Morrell.1975.JCP.62.549,Morrison.1992.JCC.13.1004,%
Pernal.2005.CPL.412.71} (EKT), $I_\text{EKT}$. For HF wave
functions, $I_\text{EKT}=-\epsilon_\text{HOMO}^\text{HF}$,
where $\epsilon_\text{HOMO}^\text{HF}$ is the eigenvalue
of the highest-occupied molecular orbital (HOMO).
For post-HF wave functions, $I_\text{EKT}$ was computed
as the largest eigenvalue of the $\mathbf{V}'$ matrix
defined in Ref.~\onlinecite{Morrison.1992.JCC.13.1004}. The
EKT ionization energies are needed to fix the constant up
to which the $v_\text{XC}$ is defined by Eqs.~\eqref{eq:vxc-2} and
\eqref{eq:vxc-3}.\cite{Ryabinkin.2015.PRL.115.083001,Cuevas-Saavedra.2015.JCP.143.244116}
This is done by shifting the potential vertically so that
$\epsilon_\text{HOMO}=-I_\text{EKT}$.

After reducing each \textit{ab initio} wave function to a
self-consistent $v_\text{XC}(\bfr)$, we evaluated the following
properties: the total noninteracting kinetic energy
\begin{equation}
  T_s = -\frac{1}{2} \sum_{i=1}^N \langle \phi_i | \nabla^2 | \phi_i \rangle,
  \label{eq:Ts}
\end{equation}
the KS exchange-correlation energy
\begin{equation}
 E_\text{XC}^\text{KS} = E_\text{XC}^\text{WF} + T_c,
\end{equation}
where
\begin{equation}
  T_c = T - T_s,
\end{equation}
and the integral
\begin{equation}
 W = \int \left[ 3\rho(\bfr) + \bfr\cdot\nabla\rho(\bfr)
   \right] v_\text{XC}(\bfr)\,d\bfr,  \label{eq:W}
\end{equation}
whose purpose will be explained shortly. The integrals in
Eqs.~\eqref{eq:T}, ~\eqref{eq:EXC-WF}, and \eqref{eq:Ts} were computed
analytically, whereas $W$ was evaluated numerically.

Strictly speaking, the quality of mRKS potentials should be judged by
their proximity to the basis-set-limit $v_\text{XC}$, but since exact
exchange-correlation potentials are rarely available, we suggest to
use weaker but feasible tests for basis-set completeness. The first
test is the integrated density error
\begin{equation}
 \Delta_\rho = \int \left| \rho^\text{KS}(\bfr)
  - \rho^\text{WF}(\bfr) \right| \,d\bfr \geq 0,   \label{eq:D-rho}
\end{equation}
where $\rho^\text{KS}(\bfr)$ is evaluated at convergence. For a given
type of wave function, $\Delta_\rho$ is uniquely determined by the
basis set used in the mRKS procedure. The premise of the test is that
$\Delta_\rho$ tends to zero as the basis set approaches completeness,
so the magnitude of $\Delta_\rho$ gives some indication of how close the
mRKS potential is to its basis-set limit. We emphasize that $\Delta_\rho$
values have entirely different meanings in KS inversion and RKS-type
methods. For instance, $\Delta_\rho\approx 0.05$ a.u.~in a KS inversion
procedure indicates that $v_\text{XC}$ is not converged, whereas in
the mRKS procedure it signals that the corresponding \textit{converged}
$v_\text{XC}$ is not yet close to the basis-set-limit potential (because
an insufficiently large basis set was used).

The second test is based on the fact that, for a given density
functional $E_\text{XC}[\rho]$ and a density $\rho(\bfr)$, the
corresponding functional derivative $v_\text{XC}(\bfr)=\delta
E_\text{XC}[\rho]/\delta\rho(\bfr)$ satisfies the virial
relation\cite{Levy.1985.PRA.32.2010}
\begin{equation}
 W = E_\text{XC}^\text{KS} + T_c,
\end{equation}
where $W$ is given by Eq.~\eqref{eq:W}.
The magnitude of the deviation
\begin{equation}
  \Delta E_\text{vir} = W - E_\text{XC}^\text{KS} - T_c
 = W - E_\text{XC}^\text{WF} - 2T_c
  \label{eq:dEvir}
\end{equation}
from zero may be taken as a measure of deviation of a trial potential from
$\delta E_\text{XC}[\rho]/\delta\rho(\bfr)$. As a quality
control test, $|\Delta E_\text{vir}|$ is more discriminating
than $\Delta_\rho$: even visually imperceptible defects of
$v_\text{XC}(\bfr)$ can result in large $\Delta E_\text{vir}$
values, as we showed previously for approximate exchange-only
potentials.\cite{Gaiduk.2008.JCP.128.204101,Staroverov.2008.JCP.129.134103,%
Ryabinkin.2013.PRL.111.013001,Kohut.2014.JCP.140.18A535}
Note that in Refs.~\citenum{Ryabinkin.2013.PRL.111.013001}
and \citenum{Kohut.2014.JCP.140.18A535} we studied KS
potentials extracted from HF wave functions as approximations
to exact-exchange potentials, for which $T_c=0$, so we defined
$\Delta E_\text{vir}=W-E_\text{X}^\text{HF}$ and evaluated
$E_\text{X}^\text{HF}$ using the KS (not HF) orbitals. This is why the
HF/UGBS values of $\Delta E_\text{vir}$ in this work are different
from those reported in Refs.\citenum{Ryabinkin.2013.PRL.111.013001}
and \citenum{Kohut.2014.JCP.140.18A535}.

\begin{table*}
\caption{Selected characteristics of various atomic wave functions and of
the exchange-correlation potentials generated from those wave functions by the RKS
and mRKS methods. $\Delta_\rho$ values are in units of electron charge, the rest are in hartrees.
The zeros not followed by decimal figures mean ``exactly zero".}
\begin{ruledtabular}
\begin{tabular}{lD{.}{.}{3.6}D{.}{.}{3.6}cD{.}{.}{3.6}D{.}{.}{1.6}D{.}{.}{1.4}D{.}{.}{3.6}D{.}{.}{1.6}D{.}{.}{1.6}}
 & & & & \multicolumn{3}{c}{RKS} & \multicolumn{3}{c}{mRKS} \\ \cline{5-7} \cline{8-10}
Basis set & \multicolumn{1}{c}{$T$} & \multicolumn{1}{c}{$E_\text{XC}^\text{WF}$} & $I_\text{EKT}$
 & \multicolumn{1}{c}{$T_s$} & \multicolumn{1}{c}{$\Delta E_\text{vir}$} 
 & \multicolumn{1}{c}{$\Delta_\rho$}
 & \multicolumn{1}{c}{$T_s$} & \multicolumn{1}{c}{$\Delta E_\text{vir}$}
 & \multicolumn{1}{c}{$\Delta_\rho$} \\ \hline
 \multicolumn{10}{l}{\quad Be, HF SCF} \\
STO-3G   & 14.844185 & -2.768067 & 0.2540 & 14.844185 & 0.003001 & 0
                                          & 14.844185 & 0.003001 & 0 \\
cc-pCVDZ & 14.571730 & -2.667161 & 0.3091 & \multicolumn{3}{c}{SCF fails to converge}
                                          & 14.583020 &  0.026191 & 0.0096 \\
cc-pCVTZ & 14.572722 & -2.666932 & 0.3093 & 14.549079 & -0.016873 & 0.0123
                                          & 14.574235 &  0.003444 & 0.0112 \\
cc-pCVQZ & 14.572929 & -2.666929 & 0.3093 & 14.582472 &  0.009668 & 0.0047
                                          & 14.572859 &  0.001138 & 0.0043 \\
UGBS     & 14.573022 & -2.666914 & 0.3093 & 14.572575 &  0.000138 & 0.0013
                                          & 14.572484 &  0.000045 & 0.0014 \\
Numerical grid\footnotemark[1] 
         & 14.573023 & -2.666914 & 0.3093 & & & & 14.572462 & 6.4\times 10^{-12} & 2.4\times 10^{-12} \\
 \multicolumn{10}{l}{\quad Be, FCI} \\
cc-pCVDZ  & 14.647784 & -2.815393 & 0.3410 & 14.634247 &  0.036533 & 0.0238
                                           & 14.584365 &  0.012058 & 0.0159 \\
cc-pCVTZ  & 14.659118 & -2.834119 & 0.3419 & 14.554768 & -0.025476 & 0.0052
                                           & 14.586875 &  0.000927 & 0.0052 \\
cc-pCVQZ\footnotemark[2]  & 14.663862 & -2.839342 & 0.3423 & 14.595868 &  0.005482 & 0.0047
                                           & 14.591807 &  0.001910 & 0.0054 \\
Basis-set limit\footnotemark[3] & 14.66736 & -2.8433  & 0.3426 & 14.5942   &    & 
                                            & 14.5942    &       &    \\
 \multicolumn{10}{l}{\quad Ne, (8,8)CASSCF} \\
3-21G    & 127.526022 & -12.331354 & 0.7418 & 127.146953 & -0.317862 & 0.0242
                                            & 127.259455 & -0.235164 & 0.0269 \\
6-31G    & 128.368644 & -12.299273 & 0.7701 & 128.074403 & -0.057903 & 0.0343
                                            & 128.207015 &  0.070006 & 0.0419 \\
6-311G   & 128.643794 & -12.310115 & 0.7889 & 128.368172 &  0.054083 & 0.0119
                                            & 128.551997 &  0.152103 & 0.0100 \\
cc-pVDZ  & 128.379168 & -12.295050 & 0.7712 & 128.091704 & -0.065663 & 0.0354
                                            & 128.217811 &  0.073355 & 0.0429 \\
cc-pVTZ  & 128.699598 & -12.313278 & 0.7972 & \multicolumn{3}{c}{SCF fails to converge}
                                            & 128.319831 & -0.440787 & 0.0154 \\
cc-pVQZ  & 128.679046 & -12.310777 & 0.8017 & 128.401283 & -0.166744 & 0.0119
                                            & 128.527364 & -0.055301 & 0.0115 \\
cc-pV5Z  & 128.681107 & -12.308871 & 0.8036 & \multicolumn{3}{c}{SCF fails to converge}
                                            & 128.544494 & -0.038485 & 0.0127 \\
cc-pV6Z  & 128.680330 & -12.308590 & 0.8038 & 129.471075 &  0.850616 & 0.0275
                                            & 128.568158 & -0.011267 & 0.0129 \\
cc-pCVDZ & 128.449457 & -12.299356 & 0.7719 & 128.585448 &  0.305618 & 0.0337
                                            & 128.447270 &  0.233908 & 0.0339 \\
cc-pCVTZ & 128.694070 & -12.315890 & 0.7978 & 128.991030 &  0.391687 & 0.0110
                                            & 128.624258 &  0.044520 & 0.0099 \\
cc-pCVQZ & 128.682785 & -12.311376 & 0.8019 & 128.639121 &  0.054117 & 0.0079
                                            & 128.583579 &  0.000298 & 0.0088 \\
cc-pCV5Z & 128.680478 & -12.308899 & 0.8036 & 128.573403 & -0.005645 & 0.0022
                                            & 128.578853 & -0.000630 & 0.0025 \\
cc-pCV6Z & 128.680103 & -12.308582 & 0.8038 & 128.579759 &  0.000595 & 0.0009
                                            & 128.579291 &  0.000100 & 0.0010 \\
UGBS     & 128.679971 & -12.308530 & 0.8039 & 128.579178 &  0.000069 & 0.0004
                                            & 128.579203 &  0.000089 & 0.0005 \\
 \multicolumn{10}{l}{\quad Ar, HF SCF} \\
STO-3G   & 512.489655 & -30.273170 & 0.4959 & 512.489655 & -0.925101 & 0 
                                            & 512.489655 & -0.925101 & 0 \\
6-31G    & 526.813061 & -30.189268 & 0.5889 & 526.566956 & -0.285033 & 0.0238
                                            & 526.598840 & -0.027741 & 0.0343 \\
6-311G   & 526.800338 & -30.186097 & 0.5901 & 528.020941 &  0.260067 & 0.0495
                                            & 526.131112 & -0.942386 & 0.0413 \\
cc-pVDZ  & 526.799649 & -30.189363 & 0.5880 & 526.499343 & -0.366185 & 0.0259
                                            & 526.552243 & -0.118648 & 0.0386 \\
cc-pVTZ  & 526.813176 & -30.186281 & 0.5901 & 526.245858 & -1.002952 & 0.0407
                                            & 526.415203 & -0.491103 & 0.0649 \\
cc-pVQZ  & 526.817051 & -30.185184 & 0.5909 & 526.517557 & -0.648849 & 0.0295
                                            & 526.548203 & -0.222111 & 0.0439 \\
cc-pV5Z  & 526.817410 & -30.185018 & 0.5910 & 526.852022 & -0.263872 & 0.0339
                                            & 526.839832 &  0.081888 & 0.0510 \\
cc-pV6Z  & 526.818234 & -30.184971 & 0.5910 & 525.371226 & -0.890411 & 0.0404
                                            & 526.994170 &  0.400168 & 0.0326 \\
cc-pCVDZ & 526.783930 & -30.189199 & 0.5880 & 519.362324 & -5.498655 & 0.1435
                                            & 526.454232 & -0.386389 & 0.0454 \\
cc-pCVTZ & 526.809131 & -30.186274 & 0.5901 & 533.394130 &  6.109779 & 0.0933
                                            & 526.383224 & -0.627448 & 0.0324 \\
cc-pCVQZ & 526.818135 & -30.185181 & 0.5909 & 531.897248 &  4.947426 & 0.0456
                                            & 526.796236 & -0.013499 & 0.0163 \\
cc-pCV5Z & 526.817374 & -30.185016 & 0.5910 & 526.701372 & -0.107472 & 0.0085
                                            & 526.812305 &  0.000905 & 0.0094 \\
cc-pCV6Z & 526.817495 & -30.184954 & 0.5910 & 526.719670 & -0.089087 & 0.0093
                                            & 526.814256 &  0.002980 & 0.0103 \\
UGBS     & 526.817656 & -30.184992 & 0.5910 & 526.816951 &  0.004847 & 0.0060
                                            & 526.811751 & -0.000259 & 0.0068 \\
Basis-set limit\footnotemark[4] & 526.817513 & & 0.5910 &          &           &   
                                             &          &          &        \\
\end{tabular}
\end{ruledtabular}
\footnotetext[1]{Numerical grid-based mRKS values from Ref.~\onlinecite{Ryabinkin.numerical.mRKS}.}
\footnotetext[2]{All $f$ and $g$ functions were removed except one $f$ with $\alpha=0.255$.}
\footnotetext[3]{Estimated exact values from Ref.~\onlinecite{Filippi.1996.Seminario.295}.}
\footnotetext[4]{Numerical HF values from Ref.~\onlinecite{Tatewaki.1994.JCP.101.4945}.}
\label{tab:1}
\end{table*}

Table~\ref{tab:1} summarizes results of RKS and mRKS calculations
for HF, CASSCF, and FCI wave functions of a few atoms.  The two
methods produce potentials with very similar $T_s$ values,
small $\Delta E_\text{vir}$, and $\Delta_\rho \sim 10^{-3}$
a.u.~when a large basis set (e.g., UGBS) is used. This is
in accord with our argument that the RKS and mRKS procedures
would be equivalent in the basis-set limit. A separate grid-based
implementation\cite{Ryabinkin.numerical.mRKS} of the mRKS procedure
for the HF wave function of Be gives $\Delta E_\text{vir}=6.4\times
10^{-12}$~$E_\text{h}$ and $\Delta_\rho=2.4\times 10^{-12}$~a.u.,
which explicitly shows that the method is KS-compliant in the basis-set limit.

\begin{figure}
\centering
\includegraphics[width=0.48\textwidth]{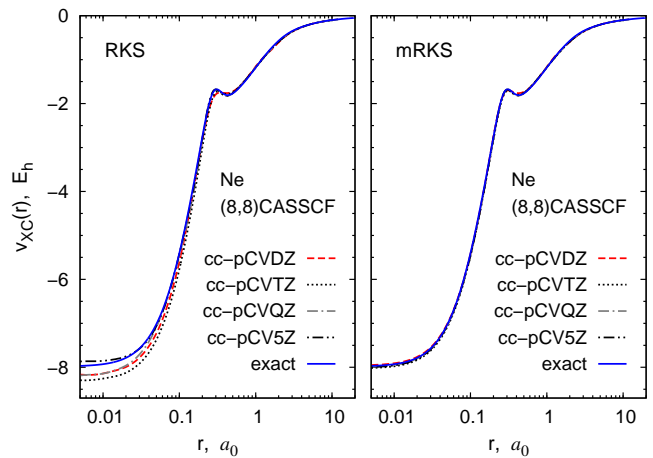}
\caption{Exchange-correlation potentials 
obtained from CASSCF/cc-pCVXZ wave functions of the Ne atom using
the RKS and mRKS methods. The exact $v_\text{XC}$ is from
Ref.~\onlinecite{Filippi.1996.Seminario.295}. The mRKS potentials
are less sensitive to basis-set incompleteness than RKS. See
Table~\ref{tab:1} for the accompanying numerical data.}
\label{fig:1}
\end{figure}

\begin{figure}
\centering
\includegraphics[width=0.48\textwidth]{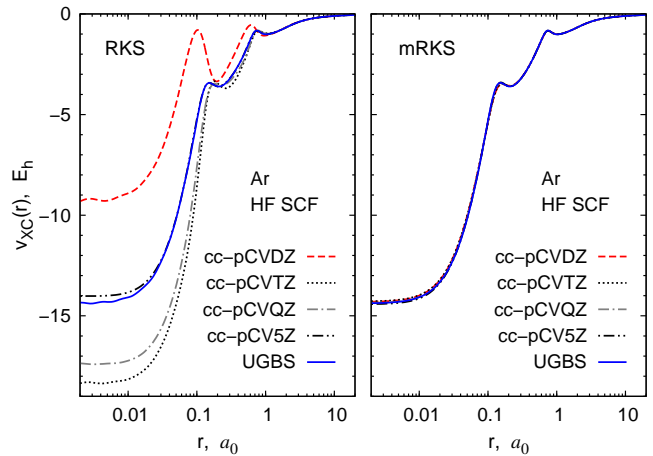}
\caption{Exchange-correlation potentials obtained from HF/cc-pCVXZ
wave functions of the Ar atom using the RKS and mRKS methods. In this
particular case, RKS potentials are accurate only if computed using
very large basis sets. See Table~\ref{tab:1} for the accompanying
numerical data.}
\label{fig:2}
\end{figure}

For small and medium basis sets, however, the original RKS method has
inconsistent performance. For instance, in the case of (8,8)CASSCF/cc-pVXZ wave
functions of the Ne atom, the RKS procedure fails to converge for the
cc-pVTZ and cc-pV5Z basis sets, and even though it converges for the
other cc-pVXZ basis sets, the results show no clear trend with respect
to basis set variations. By contrast, mRKS potentials obtained from
the same wave functions produce consistent $T_s$ values, and $|\Delta
E_\text{vir}|$ generally decreases with increasing basis-set size.
Similar observations apply to potentials generated for other atoms. In
the case of HF/cc-pCVXZ wave functions of the Ar atom, RKS potentials
for basis sets smaller than cc-pCV5Z are too high or too low near
the nucleus (Fig.~\ref{fig:2}) and have virial energy discrepancies of
up to 6 $E_\text{h}$ (Table~\ref{tab:1}). At the same time, plots of
mRKS potentials of the HF/cc-pCVXZ series are barely distinguishable
(Fig.~\ref{fig:2}). Overall, the mRKS method performs extremely well
for basis sets of any size, whereas the RKS procedure is reliable
only for large basis sets.

Figures~\ref{fig:1} and \ref{fig:2} highlight a common
feature of all RKS and mRKS potentials: they are smooth
and have no spurious oscillations that plague optimized
effective potential methods\cite{Hirata.2001.JCP.115.1635,%
Staroverov.2006.JCP.124.141103,Gorling.2008.JCP.128.104104,%
Jacob.2011.JCP.135.244102,Gidopoulos.2012.JCP.136.224109} and KS
inversion techniques that fit potentials to Gaussian-basis-set
densities.\cite{Schipper.1997.TCA.98.16,Mura.1997.JCP.106.9659,%
Silva.2012.PRA.85.032518,Kananenka.2013.JCP.139.074112,Gaiduk.2013.JCTC.9.3959}
This is because Eqs.~\eqref{eq:vxc-2} and \eqref{eq:vxc-3} contain only
terms that are well-behaved in any reasonable basis set. Note also that
for a number of potentials shown in Figs.~\ref{fig:1} and \ref{fig:2},
$|\Delta E_\text{vir}^\text{mRKS}|\ll |\Delta E_\text{vir}^\text{RKS}|$
even though $\Delta_\rho^\text{mRKS}>\Delta_\rho^\text{RKS}$.  In such
cases, the mRKS potential is visually closer to the basis-set limit,
which suggests that the virial energy discrepancy test is more sensitive
than the density error test.

\begin{figure}
\centering
\includegraphics[width=0.48\textwidth]{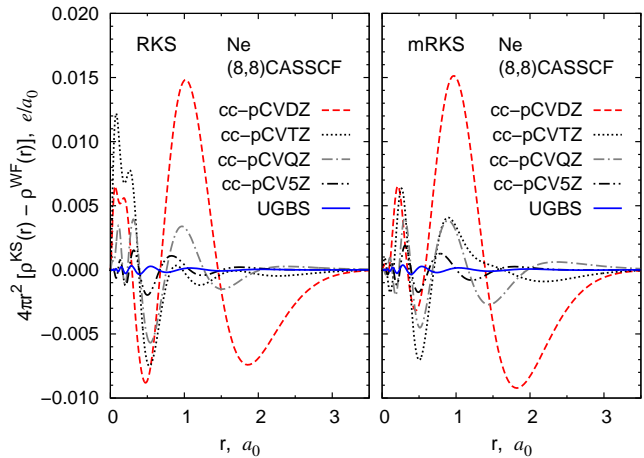}
\caption{Discrepancies between the radial KS and \textit{ab initio}
densities for the exchange-correlation potentials of Fig.~\ref{fig:1}.}
\label{fig:3}
\end{figure}

\begin{figure}
\centering
\includegraphics[width=0.48\textwidth]{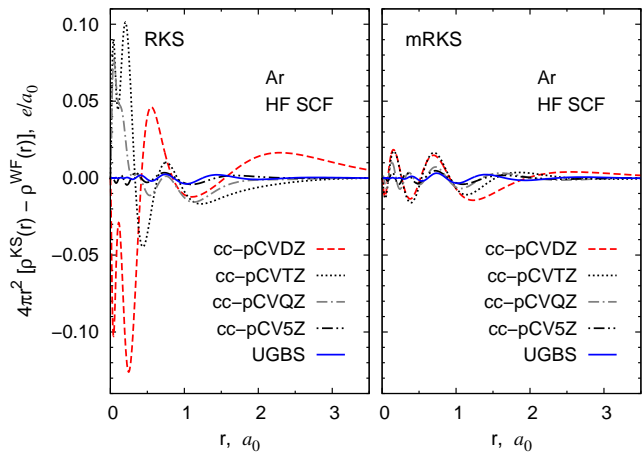}
\caption{Discrepancies between the radial KS and \textit{ab initio}
densities for the exchange-correlation potentials of Fig.~\ref{fig:2}.}
\label{fig:4}
\end{figure}

Detailed analysis of discrepancies between $\rho^\text{KS}(\bfr)$ and
$\rho^\text{WF}(\bfr)$ for the potentials shown in Figs.~\ref{fig:1}
and \ref{fig:2} furnishes another demonstration that, for large
basis sets, the RKS and mRKS procedures are practically equivalent and
produce nearly KS-compliant potentials (Figs.~\ref{fig:3} and
\ref{fig:4}). For small and medium basis sets, mRKS densities have
much smaller deviations from $\rho^\text{WF}(\bfr)$ near atomic nuclei
than do RKS densities.

RKS and mRKS calculations for HF wave functions of certain systems
exhibit a curious basis-set effect: the use of a minimal basis
sets results in $\Delta_\rho=0$ (Table~\ref{tab:1}, HF/STO-3G for
Be and Ar). This occurs when there are no virtual HF orbitals
or when no virtual orbital has the symmetry of any occupied
orbital; then the (m)RKS procedure yields occupied KS orbitals
that are unitarily transformed occupied HF orbitals, which implies
$\rho^\text{KS}=\rho^\text{HF}$. However, in such cases $\Delta
E_\text{vir}\neq 0$, meaning that the (m)RKS potential is not truly
KS-compliant.

\begin{figure}
\centering
\includegraphics[width=0.45\textwidth]{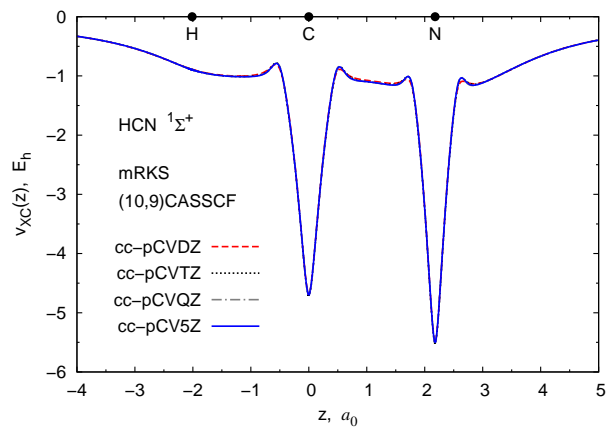}
\caption{Exchange-correlation potentials obtained by the mRKS
method from full-valence CASSCF wave functions of the HCN molecule.
See Table~\ref{tab:2} for the accompanying numerical data.}
\label{fig:5}
\end{figure}

The mRKS method also works well for molecules. To demonstrate this,
we generated exchange-correlation potentials from HF and full-valence
CASSCF wave functions of the HCN molecule using various standard
Gaussian basis sets. Here the original RKS method again failed to
converge for some basis sets, whereas the mRKS procedure converged
without difficulty in all cases (Table~\ref{tab:2}). As in the
examples involving atoms, the converged RKS and mRKS potentials
for HCN are similar and become practically identical for 
large basis sets such as cc-pCV5Z. Moreover, mRKS potentials for
HCN obtained with increasingly large basis sets of the cc-pCVXZ
($\mbox{X}=\mbox{D},\mbox{T},\mbox{Q},\mbox{5}$) series are virtually
indistinguishable by eye (Fig.~\ref{fig:5}), which shows that it is
not necessary to use large basis sets in the mRKS method to obtain
eminently reasonable potentials.

\begin{table*}
\caption{Selected characteristics of HF and full-valence CASSCF functions of the HCN
molecule at the equilibrium geometry ($r_\text{HC}=2.013a_0$, $r_\text{CN}=2.179a_0$) and of
the exchange-correlation potentials generated from those wave functions by the RKS
and mRKS methods. $\Delta_\rho$ values are in units of electron charge, the rest are in hartrees.}
\begin{ruledtabular}
\begin{tabular}{lD{.}{.}{2.6}D{.}{.}{2.6}cD{.}{.}{2.6}D{.}{.}{1.6}cD{.}{.}{2.6}D{.}{.}{1.6}c}
 & & & & \multicolumn{3}{c}{RKS} & \multicolumn{3}{c}{mRKS} \\ \cline{5-7} \cline{8-10}
Basis set & \multicolumn{1}{c}{$T$} & \multicolumn{1}{c}{$E_\text{XC}^\text{WF}$} & $I_\text{EKT}$
 & \multicolumn{1}{c}{$T_s$} & \multicolumn{1}{c}{$\Delta E_\text{vir}$} & $\Delta_\rho$
 & \multicolumn{1}{c}{$T_s$} & \multicolumn{1}{c}{$\Delta E_\text{vir}$} & $\Delta_\rho$ \\ \hline
 \multicolumn{10}{l}{\quad HCN, HF SCF} \\
6-31G*   & 92.550286 & -12.041974 & 0.4906 & 91.711955 & -0.947891 & 0.0859
                                           & 92.307987 & -0.484915 & 0.0702 \\
6-311G** & 92.742692 & -12.046387 & 0.4950 & \multicolumn{3}{c}{SCF fails to converge}
                                           & 92.734609 & -0.005396 & 0.0578 \\
cc-pCVDZ & 92.648587 & -12.046478 & 0.4925 & \multicolumn{3}{c}{SCF fails to converge}
                                           & 92.716824 &  0.105823 & 0.0501 \\
cc-pCVTZ & 92.724093 & -12.048439 & 0.4957 & 92.759771 &  0.035200 & 0.0243
                                           & 92.756065 &  0.053961 & 0.0266 \\
cc-pCVQZ & 92.728654 & -12.047784 & 0.4967 & 92.794723 &  0.064605 & 0.0122
                                           & 92.729610 &  0.008136 & 0.0128 \\
cc-pCV5Z & 92.729614 & -12.047401 & 0.4970 & 92.718948 & -0.010551 & 0.0066
                                           & 92.727317 &  0.003108 & 0.0072 \\
aug-cc-pCVDZ & 92.618227 & -12.034752 & 0.4972 &\multicolumn{3}{c}{SCF fails to converge}
                                               & 92.685523 & 0.103478 & 0.0493 \\
aug-cc-pCVTZ & 92.714623 & -12.045619 & 0.4969 & 92.734796 & 0.019105 & 0.0218
                                               & 92.743953 & 0.048228 & 0.0222 \\
aug-cc-pCVQZ & 92.726488 & -12.047072 & 0.4970 & 92.790808 & 0.061456 & 0.0113
                                               & 92.727416 & 0.007485 & 0.0114 \\
aug-cc-pCV5Z & 92.729420 & -12.047335 & 0.4970 & \multicolumn{3}{c}{SCF fails to converge}
                                               & 92.726972 & 0.002954 & 0.0062 \\
 \multicolumn{10}{l}{\quad HCN, (10,9)CASSCF} \\
6-31G*   & 92.942118 & -12.312273 & 0.5224 & 92.232843 & -0.701261 & 0.0525
                                           & 92.573544 & -0.435253 & 0.0521 \\
6-311G** & 93.093351 & -12.307472 & 0.5192 & \multicolumn{3}{c}{SCF fails to converge}
                                           & 92.974910 &  0.039466 & 0.0457 \\
cc-pVDZ  & 93.010939 & -12.305411 & 0.5168 & 92.767530 & -0.112071 & 0.0531
                                           & 92.777863 & -0.071566 & 0.0578 \\
cc-pVTZ  & 93.062905 & -12.306247 & 0.5208 & 94.792641 &  1.309037 & 0.1795
                                           & 92.800576 & -0.263447 & 0.0452 \\
cc-pVQZ  & 93.077491 & -12.307166 & 0.5213 & \multicolumn{3}{c}{SCF fails to converge}
                                           & 92.962057 & -0.006661 & 0.0347 \\
cc-pV5Z  & 93.079200 & -12.306930 & 0.5214 & 93.972821 &  0.903219 & 0.0656
                                           & 92.987779 &  0.017106 & 0.0260 \\
cc-pCVDZ & 92.988167 & -12.305592 & 0.5179 & \multicolumn{3}{c}{SCF fails to converge}
                                           & 92.886229 &  0.059178 & 0.0419 \\
cc-pCVTZ & 93.077684 & -12.307350 & 0.5212 & 93.027971 &  0.050491 & 0.0208
                                           & 93.010944 &  0.044703 & 0.0239 \\
cc-pCVQZ & 93.079442 & -12.307294 & 0.5214 & 93.046686 &  0.056174 & 0.0122
                                           & 92.988799 &  0.005245 & 0.0132 \\
cc-pCV5Z & 93.080131 & -12.307031 & 0.5214 & 92.979526 & -0.006572 & 0.0061
                                           & 92.986814 &  0.002105 & 0.0070 \\
\end{tabular}
\end{ruledtabular}
\label{tab:2}
\end{table*}

\begin{figure}
\centering
\includegraphics[width=0.45\textwidth]{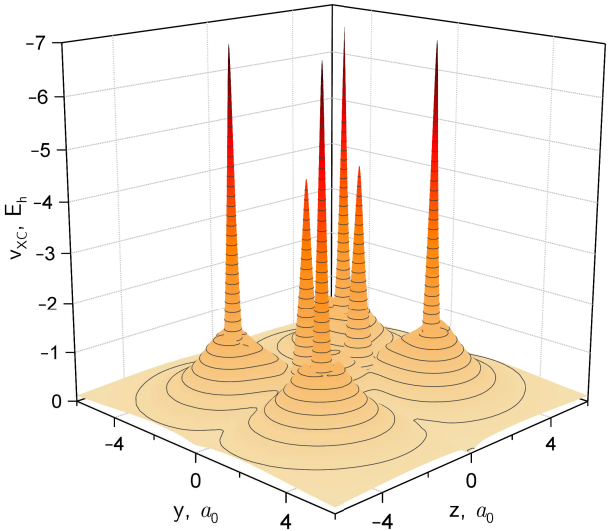}
\caption{Exchange-correlation potential of CF$_2$=CF$_2$ obtained
by the mRKS method from the HF/6-311G* wave function for the HF/6-311G*
geometry. The plot shows $v_\text{XC}$ in the molecular plane.
The $z$ axis is along the C=C bond.}
\label{fig:6}
\end{figure}

An example of an mRKS exchange-correlation potential for a polyatomic
molecule (tetrafluoroethylene) is shown in Fig.~\ref{fig:6}. RKS-type
potentials generated from HF wave functions are known to be
excellent approximations to exchange-only optimized effective
potentials.\cite{Ryabinkin.2013.PRL.111.013001,Kohut.2014.JCP.140.18A535}
The message of this figure is that molecular exchange-correlation potentials
of high quality can be effortlessly generated by the mRKS method.

\vspace*{-0.2cm}
\section{Conclusion}

We have derived Eq.~\eqref{eq:vxc-3} and showed that it
works considerably better than its predecessor, Eq.~\eqref{eq:vxc-2}, for
the purpose of generating exchange-correlation potentials from \textit{ab initio}
wave functions in
Gaussian basis sets. Equation~\eqref{eq:vxc-2} is in turn more useful
than Eq.~\eqref{eq:vxc-1}.

The transition from Eq.~\eqref{eq:vxc-1} to
Eq.~\eqref{eq:vxc-2} and then to Eq.~\eqref{eq:vxc-3} is based on the relations
\begin{equation}
 \tau_L = - \frac{1}{4} \nabla^2\rho + \tau
 = - \frac{1}{4} \nabla^2\rho + \tau_W + \tau_P
  \label{eq:taus}
\end{equation}
for each of the interacting and noninteracting systems.
For $\rho^\text{KS}=\rho^\text{WF}$, these relations imply that
\begin{equation}
 \frac{\tau_L^\text{WF}}{\rho^\text{WF}} - \frac{\tau_L^\text{KS}}{\rho^\text{KS}}
 = \frac{\tau^\text{WF}}{\rho^\text{WF}} - \frac{\tau^\text{KS}}{\rho^\text{KS}}
 = \frac{\tau_P^\text{WF}}{\rho^\text{WF}} - \frac{\tau_P^\text{KS}}{\rho^\text{KS}}.
  \label{eq:tau-diffs}
\end{equation}
Equation~\eqref{eq:taus} is always true, whereas Eq.~\eqref{eq:tau-diffs}
holds only when $\rho^\text{KS}=\rho^\text{WF}$, which in our method
happens at convergence in a complete (infinite) basis set and for
minimal-basis-set HF wave functions of certain systems. This means
that RKS-type iterations by Eqs.~\eqref{eq:vxc-1}, \eqref{eq:vxc-2},
and \eqref{eq:vxc-3} are generally not equivalent and should result in
different potentials.

In calculations using standard Gaussian basis sets, Eq.~\eqref{eq:vxc-1}
almost never converges, Eq.~\eqref{eq:vxc-2} converges for some but
not all basis sets, while Eq.~\eqref{eq:vxc-3} always converges in our
experience, at least for systems with a single-reference character. The
RKS and mRKS methods are essentially equivalent in a nearly complete
basis set, but the mRKS method is much more accurate and robust in
commonly used basis sets, making it possible to routinely generate
exchange-correlation potentials for atoms and molecules at any level of
\textit{ab initio} theory.  Therefore, we recommend the mRKS procedure
as a permanent replacement for the original RKS method.

The extensive numerical evidence presented in this work shows that
mRKS potentials generated using incomplete (finite) basis sets are
excellent approximations to the basis-set-limit $v_\text{XC}$ for
a particular type of wave function (HF, CASSCF, FCI, etc.) The mRKS
technique can also be used for construction of exchange-correlation
potentials of adiabatic time-dependent density-functional
theory.\cite{Lein.2005.PRL.94.143003,Thiele.2008.PRL.100.153004,Elliott.2012.PRA.85.052510}
Extensions of the mRKS method to spin-polarized post-HF wave functions
and to systems that are not pure-state $v$-representable remain the
subject of future work.

\vspace*{-0.6cm}
\acknowledgements

The authors thank Sviataslau Kohut for independently verifying selected
numerical results. This work was supported by the Natural Sciences and
Engineering Research Council of Canada (NSERC) through the Discovery
Grants Program (Application No.~RGPIN-2015-04814) and a Discovery
Accelerator Supplement.


\end{document}